\documentclass[twoside,accepted,11pt]{article}

\usepackage[accepted]{melba}

\usepackage{mwe} 

\usepackage{amsmath,amsfonts}

\usepackage{lscape}

\usepackage{xcolor}

\DeclareMathOperator*{\argmin}{arg\,min}

\melbaheading{2022:022}{https://www.melba-journal.org/papers/2022:022.html}{2022}{1-31}{04/2022}{09/2022}{Yu et al.}{}{}

\ShortHeadings{Validation and Generalizability of Self-Supervised Methods}{Yu et al.}
\firstpageno{1}

\title{Validation and Generalizability of Self-Supervised Image Reconstruction Methods for Undersampled MRI}

\begin{document}

\maketitle

\vspace{-20mm}

\begin{flushleft}\large
	Thomas Yu\textsuperscript{1,2},
	Tom Hilbert  \textsuperscript{1,3,4 } \footnote{Corresponding Author},
	Gian Franco Piredda\textsuperscript{1,3,4},
	Arun Joseph\textsuperscript{5,6,7},
	Gabriele Bonanno\textsuperscript{5,6,7},
	Salim Zenkhri\textsuperscript{4},
	Patrick Omoumi\textsuperscript{4}, Meritxell Bach Cuadra\textsuperscript{2,4,1},
	Erick Jorge Canales-Rodr\'iguez\textsuperscript{1},
	Tobias Kober\textsuperscript{1,3,4},
	Jean-Philippe Thiran\textsuperscript{1,2,4}
\end{flushleft}

\noindent
\begin{enumerate}
\item Signal Processing Laboratory 5 (LTS5), École Polytechnique Fédérale de Lausanne, Lausanne (EPFL), Switzerland
\item CIBM, Center for Biomedical Imaging, Switzerland 
\item Advanced Clinical Imaging Technology, Siemens Healthineers International AG, Lausanne, Switzerland
\item Department of Radiology, Lausanne University Hospital and University of Lausanne, Switzerland 
\item Advanced Clinical Imaging Technology, Siemens Healthineers International AG, Bern, Switzerland
\item Translational Imaging Center, sitem-insel, Bern, Switzerland
\item Magnetic Resonance Methodology, Institute of Diagnostic and Interventional Neuroradiology, University of Bern, Bern, Switzerland
\end{enumerate}

\bigskip

\begin{abstract}
Deep learning methods have become the state of the art for undersampled MR reconstruction. Particularly for cases where it is infeasible or impossible for ground truth, fully sampled data to be acquired, self-supervised machine learning methods for reconstruction are becoming increasingly used. However potential issues in the validation of such methods, as well as their generalizability, remain underexplored. In this paper, we investigate important aspects of the validation of self-supervised algorithms for reconstruction of undersampled MR images: quantitative evaluation of prospective reconstructions, potential differences between prospective and retrospective reconstructions, suitability of commonly used quantitative metrics, and generalizability. Two self-supervised algorithms based on self-supervised denoising and the deep image prior were investigated. These methods are compared to a least squares fitting and a compressed sensing reconstruction using in-vivo and phantom data. Their generalizability was tested with prospectively under-sampled data from experimental conditions different to the training. We show that prospective reconstructions can exhibit significant distortion relative to retrospective reconstructions/ground truth. Furthermore, pixel-wise quantitative metrics may not capture differences in perceptual quality accurately, in contrast to a perceptual metric. In addition, all methods showed potential for generalization; however, generalizability is more affected by changes in anatomy/contrast than other changes. We further showed that no-reference image metrics correspond well with human rating of image quality for studying generalizability. Finally, we showed that a well-tuned compressed sensing reconstruction and learned denoising perform similarly on all data. The datasets acquired for this paper will be made available online; see \url{https://www.melba-journal.org/papers/2022:022.html} for details.
\end{abstract}


\section{Introduction}

Since the introduction of MRI, methods for image reconstruction have evolved with acquisition acceleration and have seen great advances with parallel imaging techniques such as sensitivity encoding (SENSE) \citep{pruessmann1999sense} and generalized auto-calibrating partially parallel acquisition (GRAPPA) \citep{griswold2002generalized}. While parallel imaging reliably accelerates clinical contrasts by factors of two to three, more recent methods such as compressed sensing (CS) have achieved even higher acceleration factors \citep{lustig2007sparse}. Now, supervised deep learning methods reign as the state of the art in the reconstruction of accelerated acquisitions \citep{knoll2020deep,HAMMERNIK202025,sun2016deep}. However, these supervised methods require a non-trivial amount of fully sampled data to use as ground truth/target, which can be difficult or infeasible to obtain depending on the type of acquisition. Consequently, there has been interest in unsupervised or self-supervised, deep learning approaches which train solely on accelerated acquisitions, with no need for ground truth, fully sampled data \citep{liu2020rare,yaman2020self,heckel2019deepdecoder,akccakaya2021unsupervised}. 

However, the validation of these methods is generally done by quantitative evaluation through pixel-wise metrics on \textbf{retrospectively undersampled} acquisitions (i.e., artificial undersampling of a fully sampled dataset), sometimes accompanied by qualitative evaluation on datasets where no ground truth is available. This limitation may stem from commonly used datasets \citep{kneedataset,knoll2020fastmri} being fully sampled, as well as difficulties in acquiring datasets which contain both fully sampled and prospectively accelerated scans without motion corruption. However, this neglects quantitative evaluation of reconstructions from \textbf{prospectively undersampled} data, the clinically relevant scenario, as well as potential differences between prospective and retrospective reconstructions; furthermore, the pixel-wise metrics generally used may not correlate well with the perceptual quality of the images. This point is crucial for clinical deployment as even if different methods can be robustly ranked using retrospective data, the image quality from prospective data from the different methods may be unsuitable for clinical use. Furthermore, if these techniques will be used in future clinical routines, they likely will be subject to variations of data quality and content. For example, different surface coils, parameter differences between centers or even the use of the same sequence on different organs. Therefore, the generalizability, i.e., inference data different from the training/tuning data (e.g. in terms of field strength, sequence parameters, motion, anatomy, etc.), using prospective data is of interest, both for investigating robustness and for testing the limits of self-supervised methods. Furthermore, while prospective reconstructions are generally evaluated using qualitative rating, we evaluated the potential for using no-reference image metrics for a quantitative evaluation. 

\subsection{Contributions}
In this work, we fixed an MR sequence of interest for which extensive, clinical acquisition of fully sampled data is infeasible and conducted an extensive, realistic validation of state of the art self-supervised reconstruction methods through two novel, overarching experiments.
\begin{enumerate}
    \item In contrast to the literature, we acquired phantom data with both full sampling and prospective acceleration. This allowed us to quantitatively and qualitatively evaluate both prospective and retrospective reconstructions using both pixel-wise and perceptual metrics for fidelity to ground truth, allowing us to study them individually as well as to see any relevant differences.
    \item In contrast to the literature, we tested the generalizability of the methods using an extensive, prospectively accelerated dataset with changes in contrast, hardware, field strength, and anatomy. Furthermore, we evaluated the results both quantitatively, using no-reference image quality metrics, and qualitatively, using rating by MR scientists and a radiologist.
\end{enumerate}

\section{Theory}

\begin{figure} 
\includegraphics[width=\textwidth]{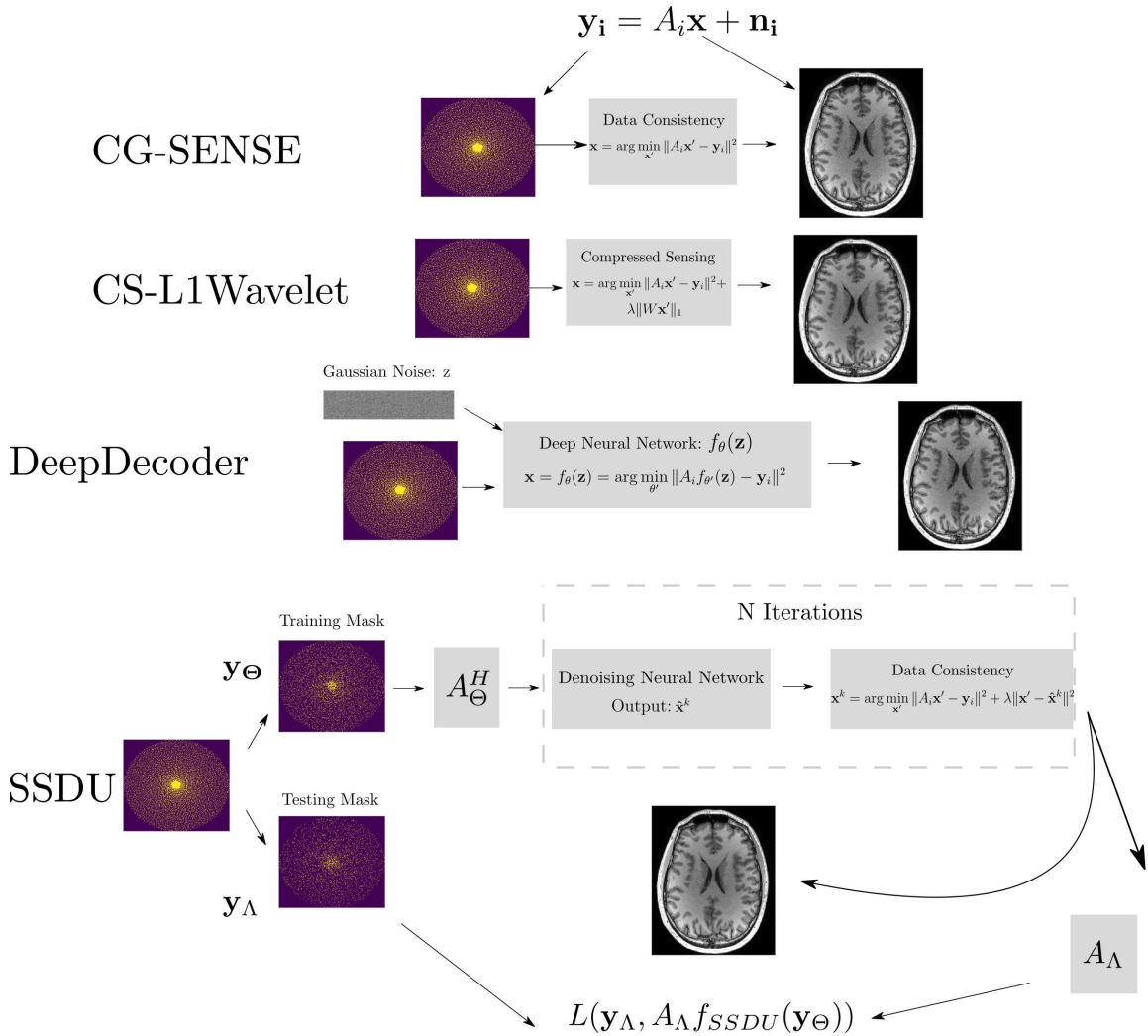}
\caption[Method Overview]{An overview of the basic formulation of the MR reconstruction inverse problem, as well as how each method in the paper solves the inverse problem.}
\label{fig:method_overview}
\end{figure}

The self-supervised, machine-learning based methods we examine in this paper rely on two powerful ideas drawn from machine learning: self-supervised denoising and restriction to the range of convolutional neural networks (CNN) as an effective prior for image reconstruction. We chose these methods for validation as these ideas have been shown to be both empirically effective and theoretically well founded, making them attractive for clinical use. In Figure \ref{fig:method_overview}, we show an overview of the different methods used in this paper. We begin with the basic inverse problem formulation of MR image reconstruction. Let $\mathbf{y}_i,\mathbf{n}_i$ denote the undersampled MR measurements and Gaussian noise respectively, from the $i$th coil element and $\mathbf{x}$ denote the underlying image. These quantities are related by:
\begin{align} \label{inverse}
    \mathbf{y}_i&= A_i\mathbf{x} + \mathbf{n}_i, \\
    A_i&=M \circ F \circ S_i
\end{align}
where $M$ is the element-wise multiplication by a mask (corresponding to the location of the undersampled measurements), $F$ denotes the Fourier transform, and $S_i$ denotes element-wise multiplication by the $i$th sensitivity map. The classical regularized reconstruction of $\mathbf{x}$ is the solution of an optimization problem
\begin{align} \label{inverse_opt}
    \mathbf{x} = \argmin_{\mathbf{x'}} D(\mathbf{x'},\mathbf{y}) + \lambda R(\mathbf{x'}), 
\end{align}
where $D(\mathbf{x},\mathbf{y})$ measures the consistency of the solution to the data (e.g. $\|A_i\mathbf{x}-\mathbf{y}_i\|^2$), $R(\mathbf{x})$ is a regularization function, which, for example, prevents overfitting to the noise, and $\lambda$ is the regularization parameter. In combination with incoherently undersampled measurements, compressed sensing reconstructions have been shown to effectively reconstruct the underlying images by setting $R(\mathbf{x})$ to encourage sparsity of $\mathbf{x}$ in a set domain \citep{lustig2007sparse}. Many state of the art deep learning methods, both supervised and unsupervised, implicitly or explicitly parametrize $R(\mathbf{x})$ with a neural network. In this work, we choose to compare two state-of-the-art self-supervised approaches which operate by orthogonal, well-founded theoretical principles with impressive empirical performance. 

\subsection{DeepDecoder}
The first self-supervised method we examine is called DeepDecoder. DeepDecoder is based on a seminal work in the machine learning literature called Deep Image Prior (DIP) \citep{ulyanov2018deep} which showed that untrained CNNs could be used to effectively solve inverse problems without ground truth.  Concretely, let $f_{\theta}$ denote a randomly initialized CNN with parameters $\theta$. Let $\mathbf{z}$ be a sample of a random, Gaussian vector.
Then DIP solves Equation \ref{inverse_opt} by
\begin{align}
    \mathbf{x}=f_{\theta}(\mathbf{z})=\argmin_{\theta'} \|A_if_{\theta'}(\mathbf{z})-\mathbf{y}_i\|^2
\end{align}
This formulation is equivalent to setting $R(\mathbf{x})$ to the indicator function with support over the range of the neural network; this assumes that the convolutional network $f_{\theta}$ itself provides a strong prior on the space of image solutions, such that only the data consistency term needs to be minimized. However, since only the noisy signal $\mathbf{y}$ is used during training, minimization can overfit the noise in the signal, depending on the inverse problem being solved (e.g. denoising, super-resolution), thus requiring early stopping \citep{ulyanov2018deep}. DeepDecoder \citep{heckel2019deepdecoder} is a CNN with a  simplified architecture (only upsampling units, pixel-wise linear combination of channels, ReLU activation, and channel-wise normalization) which is amenable to theoretical analysis and was shown to be competitive with other architectures for solving inverse problems in a DIP framework.

In \citep{heckel2020compressive}, the authors theoretically showed that for the case of image recovery from compressed sensing measurements, CNNs (in particular, CNNs with the structure of DeepDecoder) are self-regularizing with respect to noise and can simply be trained to convergence with gradient descent without early stopping or additional regularization, provided that the true, underlying image has sufficient smoothness/structure. In a knee MR example, they showed that early stopping would have only provided a marginally better solution than running to convergence. Hence, from a theoretical and practical standpoint, DeepDecoder is attractive for self-supervised reconstruction from undersampled measurements. We emphasize that DeepDecoder entails training a separate network for each separate acquisition/slice, rather than training a single network over a dataset of undersampled acquisitions.

\subsection{Self-supervised learning via data under-sampling}
 The second self-supervised method we examine is called Self-supervised learning via data under-sampling (SSDU). SSDU uses an unrolled, iterative architecture, with alternating neural network and data consistency modules, to reconstruct MR images using only undersampled measurements, with the adjoint image corresponding to the input k-space measurements as an initial guess.
 It solves Eqn \ref{inverse_opt} using an iterative, variable splitting approach where the $k$th iteration consists of
 \begin{align}
    \mathbf{\hat{x}}^k&=\text{CNN} (\mathbf{x}^{k-1}) \\
    \mathbf{x}^k&=\argmin_{\mathbf{x'}} \|A_i\mathbf{x'}-\mathbf{y}_i\|^2 + \lambda\|\mathbf{x'}-\mathbf{\hat{x}}^k\|^2.
 \end{align}
 where the superscript denotes the iteration, CNN denotes a generic CNN, and $\mathbf{\hat{x}}^k$ denotes an auxiliary variable. The regularization parameter $\lambda$ is learned during training.
 Let $f_{SSDU}$ denote the function defined by the unrolled network. In each training step of SSDU, the k-space of the data is split into two, random disjoint sets, denoted by $\mathbf{y}_\Theta$ and $\mathbf{y}_\Lambda$. $\mathbf{y}_\Theta$ is passed to the unrolled network as input. The loss function for SSDU compares the simulated k-space measurements of the corresponding image output $f_{SSDU}(\mathbf{y}_\Theta)$ to $\mathbf{y}_{\Lambda}$:
\begin{align}
    L(\mathbf{y}_\Lambda,A_{\Lambda}f_{SSDU}(\mathbf{y}_\Theta))
\end{align}
where $A_{\Lambda}$ is the measurement operator corresponding to sampling the locations of $\Lambda$, and $L$ is an equally weighted combination of the $L_1$ and $L_2$ loss. Hence, during each training step, $f_{SSDU}$ only sees information from $\mathbf{y}_\Theta$, and the loss is only computed over a disjoint set $\mathbf{y}_\Lambda$. We note that at inference time, the entire, acquired k-space measurements are given as input. While the authors of SSDU give an intuitive explanation of this approach as similar to cross validation in order to prevent overfitting to noise or learning the identity, results from the machine learning literature on blind, signal denoising can help give a theoretical explanation. 

In the Noise2Self framework \citep{batson2019noise2self}, the authors prove that a neural network can be trained to denoise a noisy signal, using solely the noisy signal for training. In the following, we describe a special case of the general theory proven in \citep{batson2019noise2self}. Let $\mathbf{y}^{\delta}=\mathbf{y}+\mathbf{n}$ denote a noisy signal, where $\mathbf{y},\mathbf{n}$ are the noise-free signal and Gaussian noise respectively. Partition $\mathbf{y}^{\delta}$ into disjoint sets, $\mathbf{y}^{\delta}_J$ and $\mathbf{y}^{\delta}_{J^C}$, where the subscript indicates restriction of the corresponding vectors to the disjoint subsets of indices $J,J^C$, with other indices being zero-filled. Then the authors showed that that a neural network (denoted as $f$) can be trained to denoise the noisy signal, using solely the noisy signal, by using the following loss function:
\begin{align} \label{noise2self}
    L(f)= \sum_{J} \mathop{\mathbb{E}} \|f_J(\mathbf{y}^{\delta}_{J^c})-\mathbf{y}^{\delta}_J\|^2 = \sum_{J} \mathop{\mathbb{E}} \|f_J(\mathbf{y}^{\delta}_{J^c})-\mathbf{y}_J\|^2 + \|\mathbf{y}^{\delta}-\mathbf{y}\|^2
\end{align}
We emphasize that the right hand side of Equation \ref{noise2self} is composed of the mean squared error between the signal predicted by the network and the ground-truth signal and a constant independent of the network.
 \textbf{Hence the Noise2Self strategy allows to minimize the error between the predicted signal and the ground truth signal with only access to the noisy signal, by iteratively giving a partition of the noisy signal as input to $f$ and computing the MSE over a disjoint partition.}  Identifying $\Theta,\Lambda$ with $J,J^c$, we can see that the training of SSDU conforms to the Noise2Self framework with the k-space measurements acting as the noisy signal, albeit with SSDU using an $L_1$ loss in addition to the $L_2$ loss. Thus, SSDU takes as input the noisy, acquired k-space measurements, and is optimized to output an image whose simulated k-space measurements are the acquired k-space measurements \textbf{without noise}. In this way, SSDU avoids overfitting to noise. This, combined with the powerful image prior from using a CNN as the neural network as well as the interleaving of the data consistency term, explains SSDU's demonstrated ability to provide denoised images which retain image sharpness, as compared to traditional methods. We can interpret SSDU as an iterative method which interleaves the application of a denoising network and a data consistency step. We note in contrast to DeepDecoder, that we can train different networks for separate acquisitions or train a single, reusable network on a dataset of undersampled acquisitions. In this paper, we do the latter. 

In conclusion, both self-supervised approaches accomplish noise robust MR reconstruction using only noisy, undersampled MR measurements;

\section{Methods}

In the following experiments, we compare four image reconstruction methods:
\begin{enumerate}
    \item \textbf{CG-SENSE}, which solves Equation \ref{inverse_opt} with no regularization using the conjugate gradient algorithm; this is a least squares fit to the acquired data similar to the description in \citep{pruessmann2001advances}. 
    \item \textbf{CS-L1Wavelet}, where we solve Equation \ref{inverse_opt} with a compressed sensing reconstruction, with $R(\mathbf{x})=\|W\mathbf{x}\|_1$, where $W$ is a wavelet transform operator. We set the regularization parameter $\lambda$ to 2.3e-4 according to a Noise2Self tuning described in the appendix.
    \item \textbf{DeepDecoder} with a depth/width of 300/10 and Gaussian input of size (10,10). 
    \item \textbf{SSDU}, where we use a U-Net \citep{ronneberger2015u} with 12 channels and 4 downsampling/upsampling layers. Training ($\Theta$) and testing ($\Lambda$) masks are randomly sampled uniformly, with a split of 60 and 40 percent respectively.
\end{enumerate}
We used Sigpy\citep{ong2019sigpy} for the computation of CS-L1Wavelet and ESPiRiT\citep{uecker2014espirit} sensitivity maps. We implemented CG-SENSE and SSDU in Pytorch \citep{paszke2019pytorch}, and we used Github implementations of DeepDecoder \footnote{https://github.com/MLI-lab/cs\_deep\_decoder} and U-Net \footnote{https://github.com/facebookresearch/fastMRI}. We used Adam \citep{kingma2014adam} to optimize both SSDU and DeepDecoder. SSDU was trained until convergence (10 epochs) with a learning rate of 0.5e-4. For each subject, DeepDecoder was optimized using the acceleration strategy in \citep{darestani2021accelerated}; a single slice for each subject is optimized to convergence (over 10,000 iterations) from a random initialization. All other slices are optimized for 1,000 iterations, initialized with the network model from this single slice. All training and inference was done on a NVIDIA Quadro RTX 8000 with 45GB of RAM. 
\subsection{Training Data and Hyperparameter Tuning}
To mimic a realistic scenario with a sequence for which fully sampled, ground truth data is difficult/infeasible to acquire, and where the training dataset is limited in size and variability, we acquired for ten healthy subjects a 5x accelerated 3D MPRAGE prototype sequence \citep{mussard2020accelerated} of the brain at 3T (MAGNETOM Prisma\textsuperscript{Fit}, Siemens Healthcare, Erlangen, Germany) using a 64ch Rx Head/Neck coil. These incoherently undersampled data were used for training/tuning the hyperparameters of all reconstruction methods. In what follows, all training/inference is done on 2D slices of both phase-encoding directions formed from performing the inverse Fourier transform along the readout direction. \textbf{We emphasize that in the absence of prior knowledge/heuristics, the hyperparameters of the methods should also be tuned in a self-supervised way, as the common method for hyperparameter tuning, i.e. using a hold-out set of data for which the ground truth is known, is not available in our scenario.} We use the Noise2Self framework, which also underlies SSDU, for selecting hyperparameters (regularization parameter of CS-L1Wavelet and the network parameters of DeepDecoder and SSDU), as it optimizes for preventing overfitting to the noise in the measurements. Details on the hyperparameter tuning can be found in the Appendix. 

\subsection{Validation using Prospectively Accelerated and Fully Sampled Data}
In our first experiment, using the aforementioned 3D MPRAGE prototype sequence used for acquiring the training/tuning data, we acquired both fully sampled and 5x prospectively accelerated scans of the following:
\begin{enumerate}
\item Siemens multi-purpose phantom E-38-19-195-K2130 filled with $MnCl_{2}\cdot4H_{2}O$ doped water
\item Assortment of fruits/vegetables (Pineapple, tomatoes, onions, brussel sprouts)
\end{enumerate}
This allowed us to reconstruct prospective, retrospective (applying the same mask as in prospective sampling on the fully sampled data), and fully-sampled images. 

No in-vivo data was used in this experiment since subject motion could bias the results. Furthermore, we used fruits/vegetables as a second phantom since they have more complex structures than a water filled container.
\subsubsection{Quantitative Assessment}
First, we qualitatively compared the results through visual inspection. Second, we quantitatively compare reconstructions to the ground truth using Peak Signal to Noise Ratio (PSNR) \citep{salomon2004data}, the Structural Similarity Index Measure (SSIM) \citep{wang2004image}, and a metric we will call the Perceptual Distance (PercDis) score. While the first two are commonly used metrics in MR image reconstruction/image reconstruction in general, the PercDis score comes from computer vision (super-resolution, style transfer, etc), where it is called the perceptual loss \citep{johnson2016perceptual}; the distance between two images is defined as the $L_1$ distance between the respective induced features from intermediate layers of a pretrained image classification network. The scores of center cropped slices, along the read-out direction, are averaged for the final score. 

\subsection{Generalizability of Self-Supervised Reconstruction Methods}
In our second experiment, we examined the generalizability of the reconstruction methods. To that end, we scanned three, healthy subjects with the following prospectively accelerated sequences(anatomy):
\begin{enumerate}
    \item 1.5T MPRAGE (Brain)
    \item 3T MPRAGE (Brain)
    \item 7T MPRAGE (Brain)
    \item 3T MPRAGE with 1Tx/20Rx Coil (Brain)
    \item 3T MPRAGE with Subject Motion  (Brain)
    \item 3T MPRAGE with Different Parameters (Brain)
    \item 3T, $T_1$ SPACE (Brain)
    \item 3T, $T_2$ FLAIR SPACE (Brain)
    \item 3T, PD SPACE (Knee)
    \item 3T, $T_2$ SPACE (Knee)
\end{enumerate}

The brain scans at 1.5T, 3T and 7T (MAGNETOM Sola, Vida, and Terra, Siemens Healthcare, Erlangen, Germany) were done using a 1Tx/20Rx, 1Tx/64Rx (unless otherwise stated), and 8pTx/32Rx (Nova Medical, Wilmington, MA, USA) head coil, respectively. The knee scans at 3T were done with a 1Tx/18Rx coil. All detailed sequence parameters can be found in the Appendix in Table \ref{tab:seqparam}.

As ground truth data is not available since motion would render quantitative comparison difficult due to blurring from image co-registration, we evaluated the reconstructions from the above data quantitatively through no-reference image quality metrics and qualitatively through rating by four MR scientists and a radiologist. In total, 120 reconstructions (40 per subject) were evaluated. 
\subsubsection{No-Reference Image Metrics}
No-reference image quality metrics quantify the quality of a given image (i.e. blurriness, noise) using only its statistical features in a way that correlates with the perceptual quality of a human observer. They have been shown to potentially be useful for MR/medical image evaluation without ground truth \citep{woodard2006no,zhang2018can}; we use the following three metrics: a metric used originally for assessing the quality of JPEG-compressed images which we call NRJPEG \citep{wang2002no}, the Blind/Referenceless Image Spatial Quality Evaluator (BRISQUE) \citep{mittal2011blind}, and Perception based Image Quality Evaluator (PIQE) \citep{venkatanath2015blind}. BRISQUE and PIQE have also been used in other image reconstruction challenges where the ground truth is not available, such as super-resolution \citep{SRchallenge}. The metrics were calculated for the central 100 slices (along the read-out direction) of each reconstruction. 
\subsubsection{Human Quality Rating}
The human quality rating was done according to \citep{hammernik2018learning} by four experienced MR scientists and a radiologist. Using a 4-point ordinal scale, reconstructed images were evaluated for sharpness (1: no blurring, 2: mild blurring, 3: moderate blurring, 4: severe blurring), SNR (1: excellent, 2: good, 3: fair, 4: poor), presence of aliasing artifacts (1: none, 2: mild, 3: moderate, 4: severe) and overall image quality (1: excellent, 2: good, 3: fair, 4: poor). Raters were blinded to the reconstruction method.

\subsection{Statistical Significance}
For all quantitative metrics/ratings, we use the Wilcoxon signed rank test with significance level $\frac{0.05}{6}$ (Bonferroni correction with 6 pair-wise comparisons among the 4 methods) to determine statistical significance.

\section{Results}
In general, perceptually, CG-SENSE produces noisy but sharp images since it is not regularized. DeepDecoder produces smoother reconstructions with spatially varying noise behavior and sharpness, e.g Figure \ref{fig:phantom} (yellow arrows). CS-L1Wavelet and SSDU produce similar images, smoother than those of CG-SENSE with comparable sharpness; however, CS-L1Wavelet exhibits more artifacts, e.g Figure \ref{fig:phantom} (red arrows).

\subsection{Validation Using Prospectively Accelerated and Fully Sampled Data}

\begin{figure} 
\includegraphics[width=\textwidth]{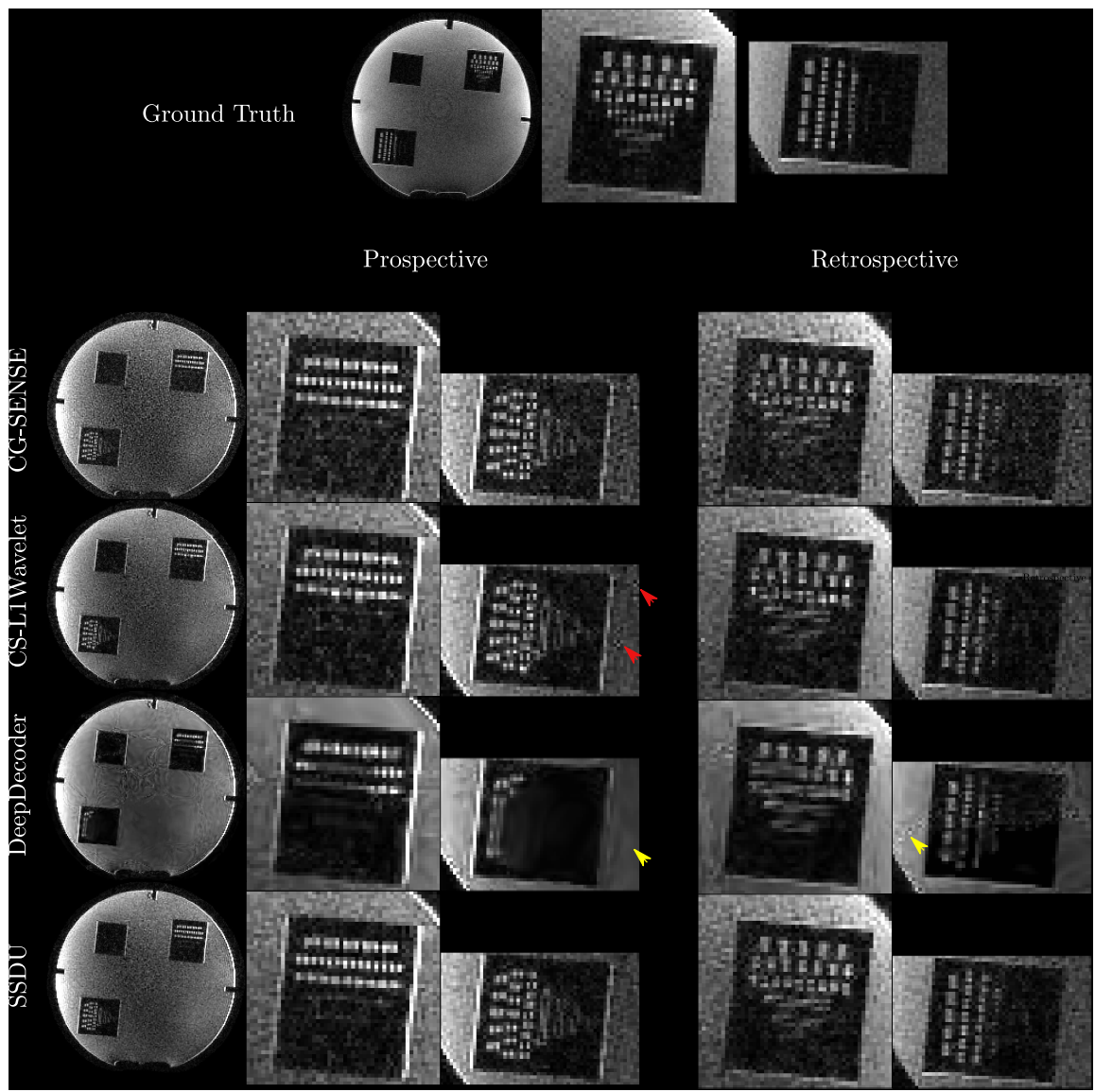}
\caption[Prospective/Retrospective Reconstructions of a Multi-Purpose Phantom]{Ground truth images and reconstructed images using prospectively and retrospectively accelerated data from the multi-purpose phantom, scanned with a MPRAGE sequence at 3T. Reconstructions from prospectively accelerated data are distorted (see closeups) relative to the ground truth/retrospective reconstructions. DeepDecoder exhibits spatially varying smoothness/distortion (see yellow arrows) relative to CS-L1Wavelet and SSDU which have similar scores/appearance, although CS-L1Wavelet has more artifacts (see red arrows). CG-SENSE produces noisy but sharp reconstructions, while CS-L1Wavelet and SSDU reduce noise but preserve sharpness relative to CG-SENSE.
}
\label{fig:phantom}
\end{figure}

\begin{figure} 
\includegraphics[width=\textwidth]{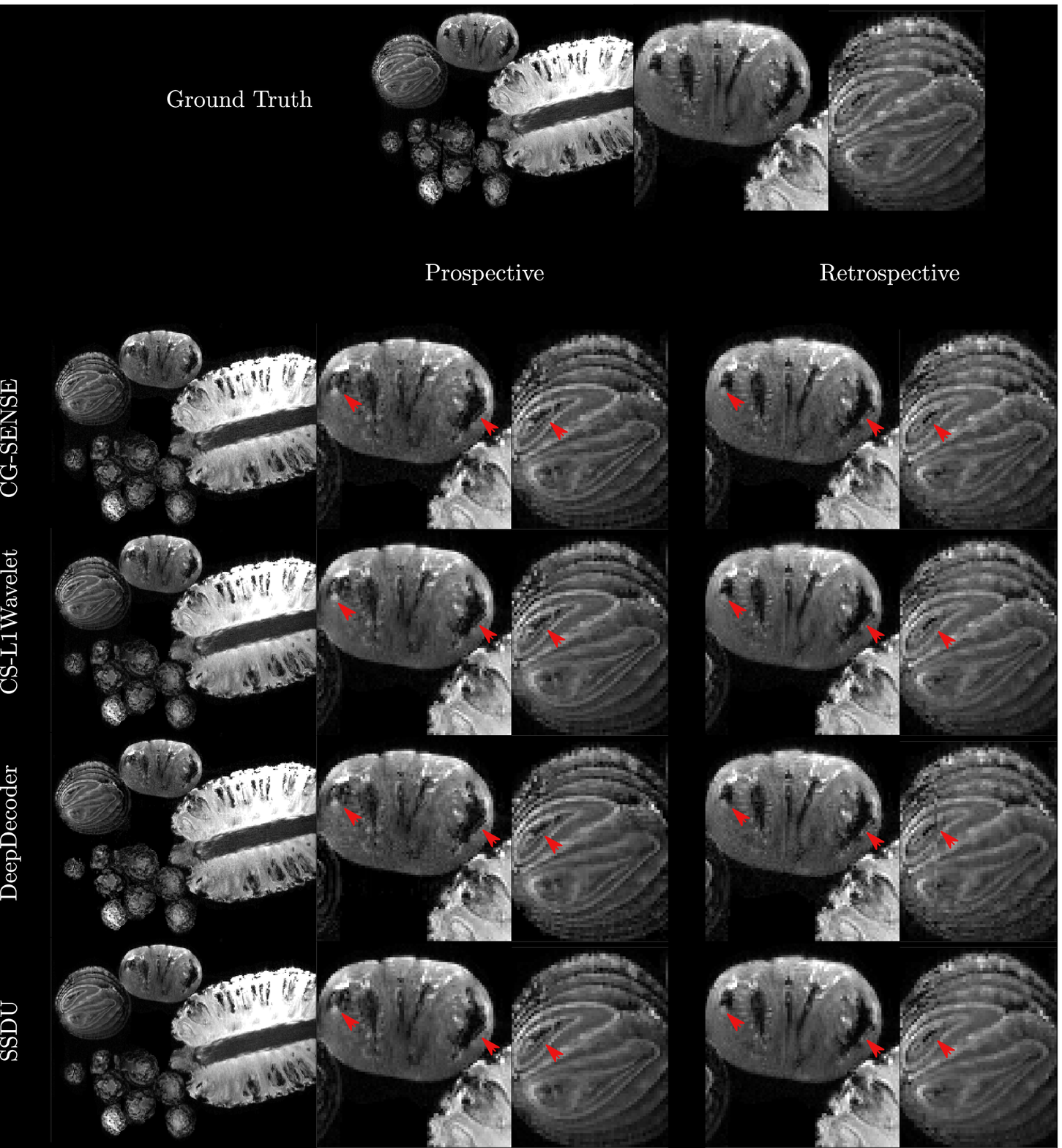}
\caption[MPRAGE: Prospective/Retrospective Reconstructions of Fruits/Vegetables]{Ground truth images as well as reconstructed images using prospectively and retrospectively accelerated data from the fruits/vegetables, scanned with a MPRAGE sequence at 3T. Reconstructions from prospectively accelerated data are distorted in hypointense regions (see closeup/red arrows) relative to the ground truth/retrospective reconstructions. Qualitatively, the main difference between the methods are between CS-L1Wavelet/SSDU and CG-SENSE/DeepDecoder, where the former group is smoother than the latter.
}
\label{fig:fruits_et_legumes}
\end{figure}

 In Fig. \ref{fig:phantom} and Fig. \ref{fig:fruits_et_legumes}, we can see spatial distortions of hyper/hypo-intense features in the prospective reconstructions and changes in contrast in comparison to the ground truth reconstruction; this distortion is not present in the retrospective reconstructions; however, they are similar across all reconstruction methods. 
 
Retrospective reconstructions have significantly higher mean scores for all metrics in comparison to the prospective reconstructions in both acquisitions (see Table \ref{tab:psnr_ssim}).
 
  Comparing the methods, in the phantom, the prospective/retrospective reconstructions of DeepDecoder have the highest pixel-wise fidelity to the ground truth with a mean PSNR of (18.67/23.44) and SSIM of (0.49/0.52); however, qualitatively, it has more spatially varying oversmoothing than those of CS-L1Wavelet and SSDU. SSDU and CS-L1Wavelet perform similarly, with the highest qualitative similarity to the ground truth, with SSDU having a higher mean PSNR overall (17.79/21.95). In contrast to the PSNR/SSIM results, with the PercDis score, SSDU has the highest fidelity to the ground truth (0.63/0.61). 
  
  Qualitatively and quantitatively (with PSNR and SSIM), the differences between the methods are much less in the fruits/vegetables. The main qualitative difference is the greater denoising capabilities of SSDU and CS-L1Wavelet in comparison to CG-SENSE and DeepDecoder. Quantitatively, there are only minor differences between the methods with respect to PSNR and SSIM. In contrast, the PercDis scores clearly indicate that CS-L1Wavelet and SSDU (with similar scores) are perceptually more similar to the ground truth than CG-SENSE and DeepDecoder (with similar scores).
  \begin{table}
\begin{tabular}{lllll}
\cline{1-1}
\multicolumn{1}{|l|}{\textbf{PSNR} $\uparrow$}            & \textbf{Phantom}                          &                                    & \textbf{Fruits/Vegetables}                &                                    \\ \hline
\multicolumn{1}{|l|}{($\mu,\sigma$)}       & \multicolumn{1}{l|}{Prospective} & \multicolumn{1}{l|}{Retrospective} & \multicolumn{1}{l|}{Prospective} & \multicolumn{1}{l|}{Retrospective} \\ \hline
\multicolumn{1}{|l|}{CG-SENSE}        & (13.54,9.69)                     & (16.82,12.17)                      & (33.4,4.86)                      & (39.3,5.73)                        \\ \cline{1-1}
\multicolumn{1}{|l|}{CS-L1Wavelet}       & (16.0,8.36)                      & (20.62,11.65)                      & (33.59,3.83)                     & (38.88,4.44)                       \\ \cline{1-1}
\multicolumn{1}{|l|}{DeepDecoder}     & (18.67,6.65)                     & (23.44,10.66)                      & (33.65,2.86)                     & (38.25,4.42)                       \\ \cline{1-1}
\multicolumn{1}{|l|}{SSDU}            & (17.79,6.9)                      & (21.95,10.09)                      & (33.88,3.75)                     & (39.49,4.27)                       \\ \cline{1-1}
                                      &                                  &                                    &                                  &                                    \\ \cline{1-1}
\multicolumn{1}{|l|}{\textbf{SSIM} $\uparrow$}            &                           &                                    &                 &                                    \\ \hline
\multicolumn{1}{|l|}{($\mu,\sigma$)}       & \multicolumn{1}{l|}{Prospective} & \multicolumn{1}{l|}{Retrospective} & \multicolumn{1}{l|}{Prospective} & \multicolumn{1}{l|}{Retrospective} \\ \hline
\multicolumn{1}{|l|}{CG-SENSE}        & (0.35,0.18)                      & (0.42,0.23)                        & (0.92,0.09)                      & (0.95,0.09)                        \\ \cline{1-1}
\multicolumn{1}{|l|}{CS-L1Wavelet}       & (0.4,0.21)                       & (0.47,0.27)                        & (0.93,0.08)                      & (0.96,0.08)                        \\ \cline{1-1}
\multicolumn{1}{|l|}{DeepDecoder}     & (0.49,0.22)                      & (0.52,0.28)                        & (0.93,0.04)                      & (0.95,0.08)                        \\ \cline{1-1}
\multicolumn{1}{|l|}{SSDU}            & (0.41,0.22)                      & (0.47,0.27)                        & (0.93,0.08)                      & (0.96,0.08)                        \\ \cline{1-1}
                                      &                                  &                                    &                                  &                                    \\ \cline{1-1}
\multicolumn{1}{|l|}{\textbf{PercDis} $\downarrow$} &                           &                                    &                 &                                    \\ \hline
\multicolumn{1}{|l|}{($\mu,\sigma$)}     & \multicolumn{1}{l|}{Prospective} & \multicolumn{1}{l|}{Retrospective} & \multicolumn{1}{l|}{Prospective} & \multicolumn{1}{l|}{Retrospective} \\ \hline
\multicolumn{1}{|l|}{CG-SENSE}        & (1.05,0.08)                      & (1.02,0.06)                        & (0.45,0.16)                      & (0.29,0.09)                        \\ \cline{1-1}
\multicolumn{1}{|l|}{CS-L1Wavelet}       & (0.84,0.08)                      & (0.79,0.05)                        & (0.41,0.17)                      & (0.25,0.09)                        \\ \cline{1-1}
\multicolumn{1}{|l|}{DeepDecoder}     & (0.68,0.15)                      & (0.64,0.09)                        & (0.44,0.19)                      & (0.3,0.11)                         \\ \cline{1-1}
\multicolumn{1}{|l|}{SSDU}            & (0.63,0.13)                      & (0.61,0.1)                         & (0.42,0.16)                      & (0.26,0.09)                        \\ \cline{1-1}
\end{tabular}
\caption[PSNR/SSIM/PercDis with respect to Ground Truth]{Mean and standard deviation of PSNR/SSIM/PercDis scores of the reconstructions with respect to the ground truth for the phantom and the fruit/vegetables; arrows beside each metric denote whether higher or lower values are better. PSNR/SSIM/PercDis were calculated over all the slices in the read-out direction with center cropping. We found statistically significant differences between each method for each metric \textbf{other than} (CS-L1Wavelet vs SDDU, Retrospective SSIM, Phantom) and (CG-SENSE vs SSDU, Retrospective PSNR, Fruits/Vegetables) Note that while with respect to PSNR/SSIM, DeepDecoder performs the best in the phantom, and all methods perform similarly in Fruits/Vegetables. In contrast, with respect to the PercDis score, SSDU performs the best in both cases by larger relative margins than with PSNR/SSIM.}
\label{tab:psnr_ssim}
\end{table}

\subsection{Generalizability}
\begin{figure} 
\includegraphics[width=\textwidth]{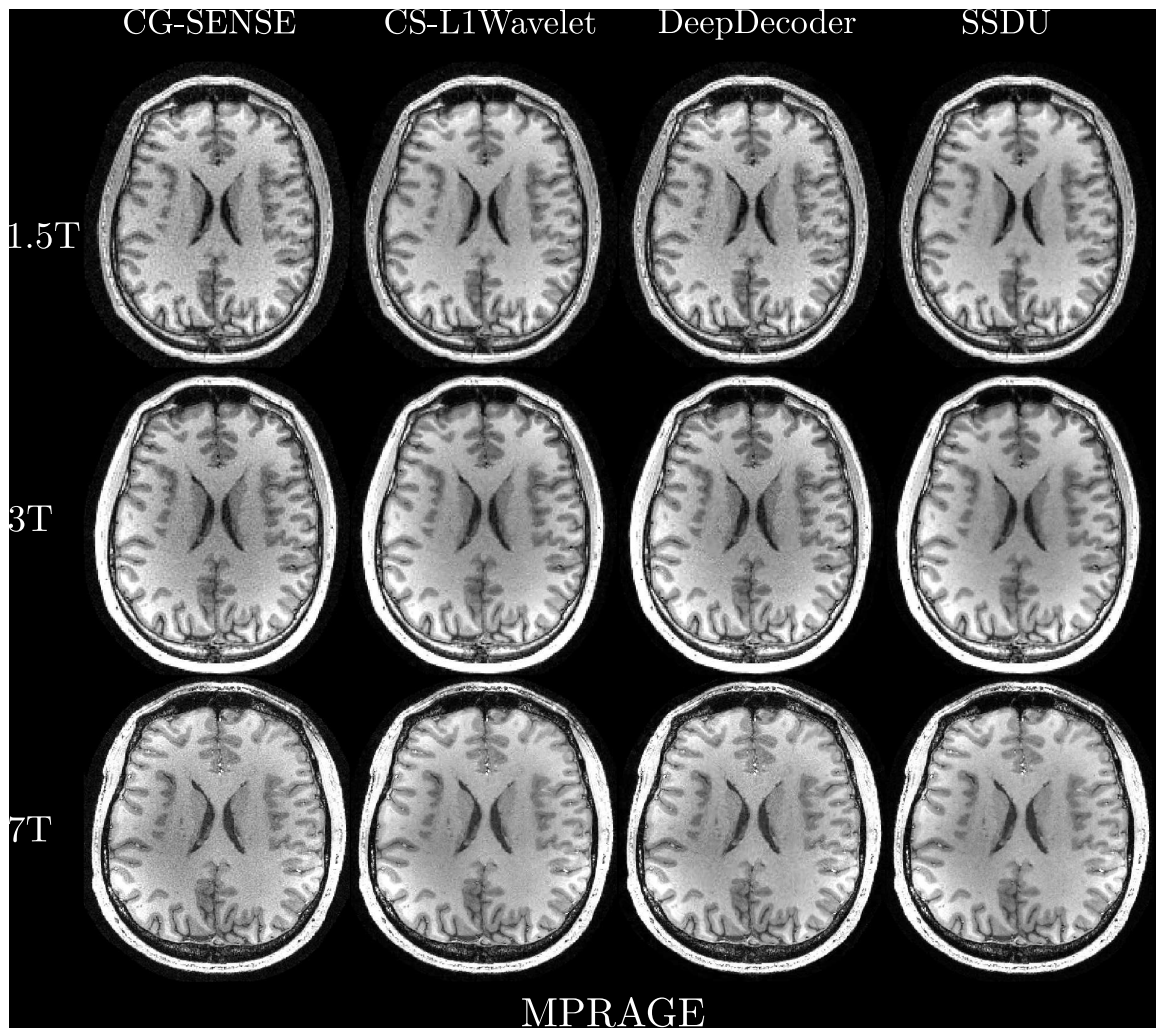}
\caption[MPRAGE: Axial Brain Slices]{Axial slices from prospective reconstructions of MPRAGE scans of the brain at different field strengths. \textbf{Images are not co-registered}; The interpolation of image co-registration introduces blurring and thus was omitted. We chose slices at similar locations for visualization. CG-SENSE produces noisy but sharp reconstructions, and DeepDecoder produces smoother reconstructions with spatially varying noise and oversmoothing. CS-L1Wavelet and SSDU produce similarly smooth/sharp reconstructions. 
 }
\label{fig:multiple_field_strength}
\end{figure}

\begin{figure} 
\includegraphics[scale=1.5]{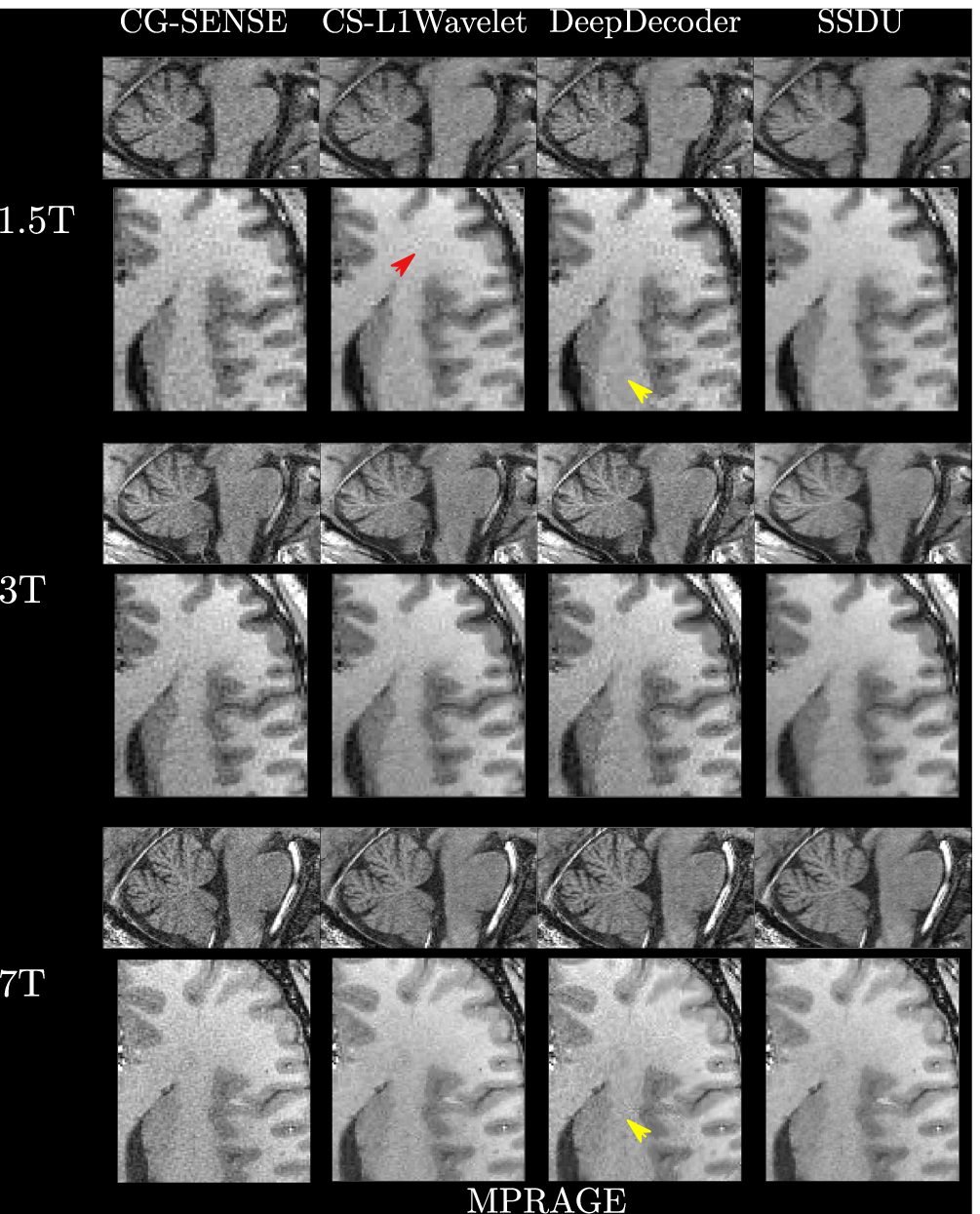}
\caption[MPRAGE: Closeups of the cerebellum and an axial slice]{Closeups of the prospective reconstructions of MPRAGE scans of the brain at different field strengths; we show closeups of the cerebellum in a sagittal view as well as the left frontal lobe in an axial view. In the axial closeups, the spatially varying smoothness of DeepDecoder is apparent (yellow arrows); furthermore, wavelet artifacts of CS-L1Wavelet can be seen in, for example, the axial closeup at 1.5T (red arrow). In general, we can see that all methods improve in sharpness (as can be seen from the closeups of the corpus callosum) with increasing field strength.}
\label{fig:multiple_field_strength_closeup}
\end{figure}

Figures \ref{fig:multiple_field_strength},~\ref{fig:multiple_field_strength_closeup} show axial MPRAGE brain slices at the different field strengths and corresponding closeups of the cerebellum and the left frontal lobe. Figure \ref{fig:knee} shows a sagittal PD knee slice (3T) with closeups of articular cartilage interfaces in sagittal (femur) and axial (patella) views. These show the generalizability of the methods to different magnetic field strengths as well as changes in anatomy and contrast. Example reconstructions for the other sequences can be found in the Appendix \ref{S-fig:mprage_variations},~\ref{S-fig:space_variations}. 

\subsubsection{Perceptual Evaluation}
\begin{figure} 
\includegraphics[scale=0.8]{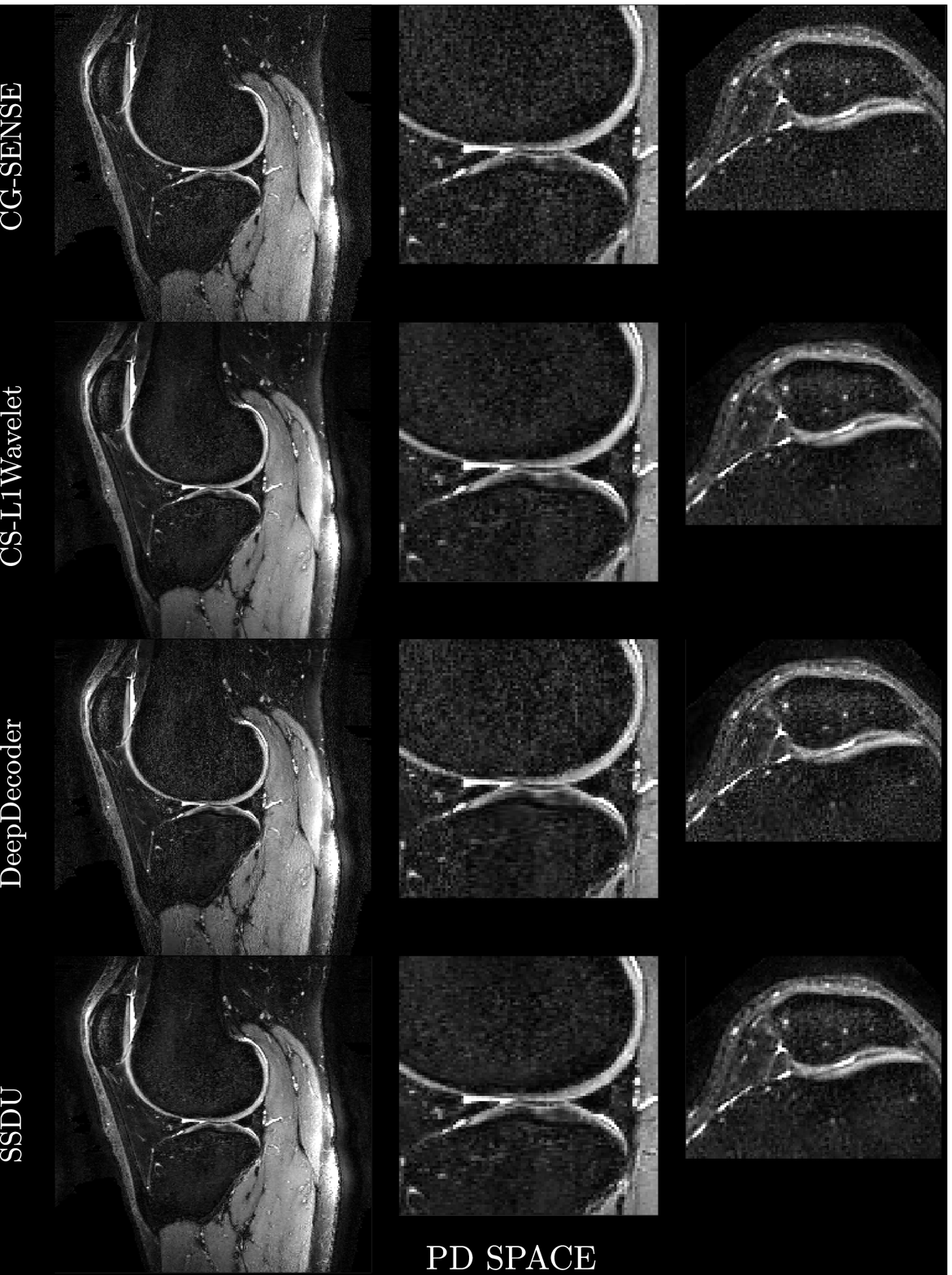}
\caption[PD SPACE: Sagittal slice of the knee with sagittal/axial closeups]{Prospective reconstructions from PD SPACE scans of the knee, where we show a sagittal slice as well as closeups on the articular cartilage interface in sagittal (femur) and axial (patella) views. Qualitatively, the main differences are between CS-L1Wavelet/SSDU and CG-SENSE/DeepDecoder, where the former group removes noise better than the latter.}
\label{fig:knee}
\end{figure}

Qualitatively, we can see from Figures \ref{fig:multiple_field_strength},~\ref{fig:multiple_field_strength_closeup},~\ref{fig:knee} that all methods are able to generalize well (in the sense of approximately preserving performance/appearance on dataset used for training/tuning) to changing field strengths, anatomy, and contrast, although changing anatomy clearly worsened absolute image quality as compared to changing field strength. DeepDecoder preserves its spatially varying smoothing/artifacts, and SSDU/CS-L1Wavelet are able to produce images with less noise and comparable sharpness to CG-SENSE, although CS-L1Wavelet exhibits more artifacts. As expected, the perceptual quality of all methods increase with increasing field strength due to higher spatial resolution. Differences between the methods are less pronounced in the knee scan although overall image quality is worse. 
\subsubsection{No-reference Image Quality Metrics}
In the first row of Figure \ref{fig:barplots}, we show a bar plot of the scores for the no-reference image quality metrics averaged over all sequences and subjects. 
In general, CS-L1Wavelet and SSDU have the highest (by a small margin) mean NRJPEG score (10.54/10.39) and lowest, mean BRISQUE (29.35/28.06) and PIQE (25.56/22.87) scores, indicating better image quality in comparison to CG-SENSE and DeepDecoder. 
\subsubsection{Human Ratings}

\begin{figure} 
\includegraphics[width=\textwidth]{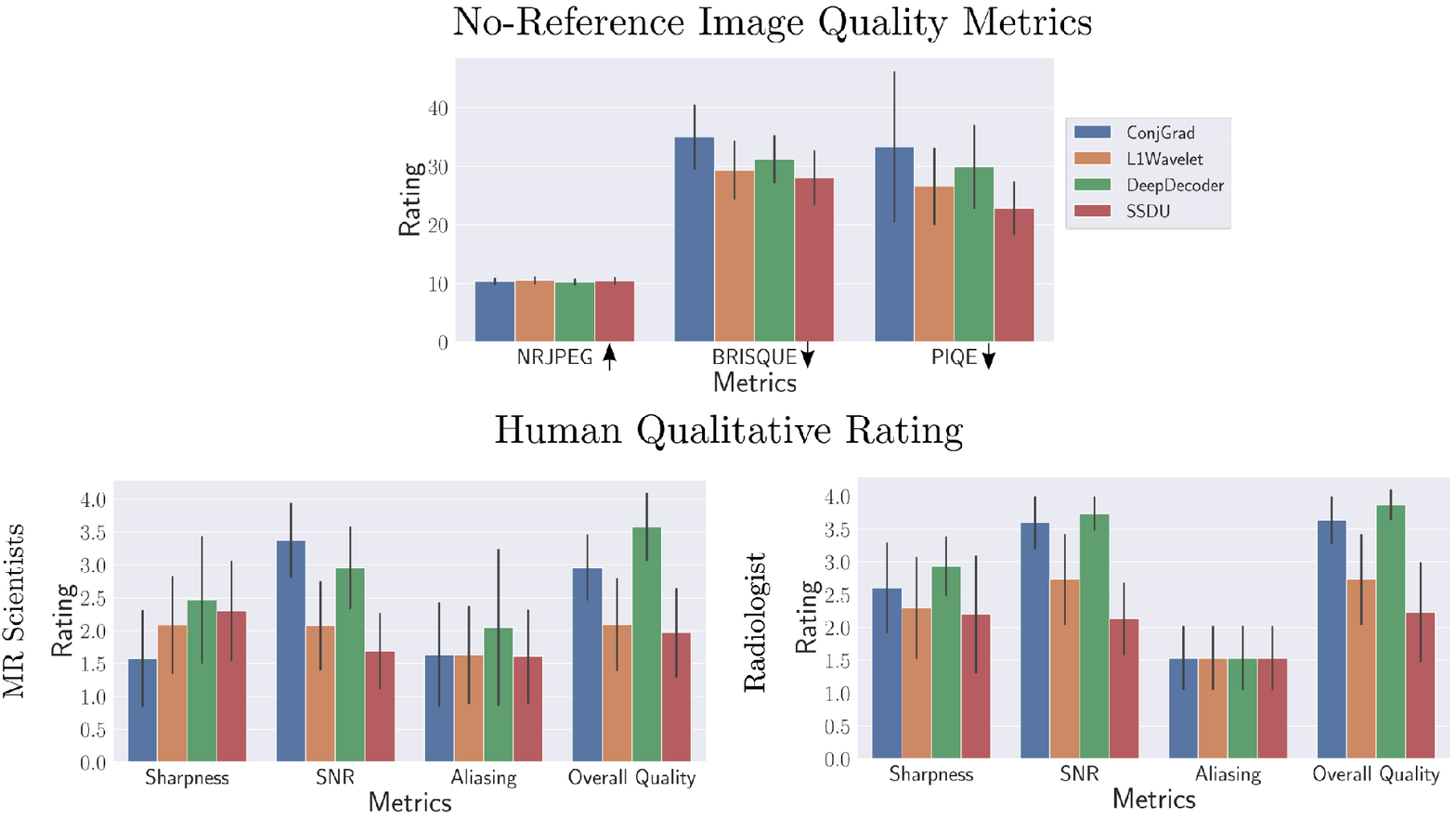}
\caption[Barplots of No-reference image metrics and human ratings]{Barplot of the no-reference image metrics averaged over all the subjects/different sequences in the generalizability study (top row). The arrow next to each metric indicates whether higher/lower scores are better. Barplots of the qualitative rating done by the MR scientists (pooled together) and the radiologist respectively (bottom row). We found statistically significant differences between all methods with respect to the no-reference image metrics. With respect to the MR scientists the following differences \textbf{were not} statistically significant: (DeepDecoder,CS-L1Wavelet,Sharpness),(DeepDecoder,SSDU,Sharpness), all of the aliasing comparisons, and (CS-L1Wavelet, SSDU, Overall Quality). All other comparisons were found to be statistically significant. With respect to the radiologist, all sharpness and aliasing comparisons were found to \textbf{not be} statistically significant. In the SNR comparisons, only (CG-SENSE/DeepDecoder vs. SSDU) were found to \textbf{be} statistically significant. In the overall quality comparisons, only (CG-SENSE vs DeepDecoder) and (CS-L1Wavelet vs. SSDU) were found to \textbf{not be} statistically significant. Overall, the no reference image metrics and human rating agree that CS-L1Wavelet/SSDU exhibit better overall image quality than DeepDecoder/CG-SENSE.

}
\label{fig:barplots}
\end{figure}
In the second row of Figure \ref{fig:barplots}, we show bar plots of the scores from the MR scientists and the radiologist; we pooled the scores of the MR scientists. We see that MR scientists and the radiologist generally agree for evaluating SNR, aliasing, and overall quality, rating CS-L1Wavelet/SSDU as being better than or the same as CG-SENSE/DeepDecoder. We recall that lower ratings correspond to better quality. MR scientists rated CS-L1Wavelet/SSDU with a mean overall quality of (2.09/1.97) as compared to CG-SENSE/DeepDecoder with (2.96/3.57). The radiologist rated CS-L1Wavelet/SSDU with a mean overall quality of (2.73/2.23) as compared to CG-SENSE/DeepDecoder with (3.63/3.87). We note that for both sets of raters, the difference between CS-L1Wavelet and SSDU in overall image quality was found to not be statistically significant. Furthermore, when we restrict our analysis to the average score change between the subgroup of changes in field strength vs. the subgroup of PD Knee/$T_2$ Knee scans, the overall image quality rating of CG-SENSE/CS-L1Wavelet/DeepDecoder/SSDU all worsen in the knee scans for the MR scientists, with increases of 0.26,0.40,0.11, and 0.79 respectively. In contrast, for the radiologist, this shift results in changes of -0.33,0.33,-0.16, and 0.83 respectively, indicating that only CS-L1Wavelet and SSDU worsened.
\section{Discussion}

In contrast to the previous literature, this work critically examines the validation and generalizability of self-supervised algorithms for undersampled MRI reconstruction through novel experiments with a focus on prospective reconstructions, the clinically relevant scenario. To this end, we analyze results from acquiring both fully-sampled and prospectively accelerated data on two phantoms and prospectively accelerated, in-vivo data over a wide variety of different sequences.

\subsection{Validation using Prospectively Accelerated and Fully Sampled Data}
Concerns about the differences between prospective and retrospective reconstructions were also raised in \citep{muckley2021results}, in the context of end-to-end, supervised methods for parallel MR image reconstruction. In particular, they noted that retrospective undersampling neglects potential differences in signal relaxation across echo trains, and verification should be performed before clinical use. 
From our results using both fully sampled and prospectively accelerated data, it is clear that for the 3D MPRAGE sequence, prospective vs. retrospective reconstructions can differ meaningfully, with retrospective reconstructions having greater fidelity to the fully sampled reconstruction; prospective reconstructions exhibit spatial distortions and local changes in contrast, with respect to the ground truth. This is despite the methods being tuned/trained on prospectively accelerated data; hence, this can be attributed to the differences in the prospectively vs. retrospectively sampled k-space data, potentially due to the different gradient patterns used in the sequences. This difference is relevant both for self-supervised and supervised machine learning methods; indeed, end-to-end, supervised methods which are trained on retrospective data may yield even greater distortion than self-supervised methods when prospective data is used for inference. However, the performance ranking of the different methods was the same in both prospective and retrospective reconstructions. Therefore, retrospective image quality cannot necessarily be taken as a reliable proxy for prospective image quality; however, they can be used to show differences between methods. 

The quantitative results in the phantom show how ranking by PSNR and SSIM can be misleading, as images that are perceptually/qualitatively more similar to the ground truth (SSDU,CS-L1Wavelet) can have significantly worse or almost identical mean PSNR/SSIM scores than images which are less qualitatively similar (CG-SENSE,DeepDecoder). In contrast, ranking with the PercDis score, which measures distances between the feature activations within a pretrained classification network of the images rather than the images themselves, better matches with the perceptual quality of the images, showing that SSDU or SSDU/CS-L1Wavelet are better, by a significant margin (relatively with respect to the same differences in PSNR/SSIM), than the other methods. The PercDis score or perceptual loss \citep{johnson2016perceptual} was created precisely because they found this metric better suited for measuring perceptual similarity than PSNR/SSIM. This apparent tradeoff between PSNR/SSIM and perceptual similarity is well-known in the computer vision community, where it is called the perception-distortion tradeoff \citep{blau2018perception}. This concept has also recently been explored in MR; in \citep{adamson2021ssfd}, the authors train an in-painting network on the Fastmri dataset, and use the features of intermediate layers for quantitative evaluation, producing a perceptual distance tailored for MR images. In \citep{Wang2019HighFidelityRW}, the authors propose a new reconstruction method which uses distances in feature space (trained from ground truth MR reconstructions) to better recover textures/perceptual appearance than using just pixel-wise metrics.   

\subsection{Generalizability}
We note that as our generalizability study is conducted on prospective reconstructions, which we showed can exhibit distortions relative to fully-sampled reconstructions, it cannot be considered as clinical validation; however, as all methods are affected the same way, this study still can give a good idea of how well each method generalizes. 
While one might conjecture that generalizability is less of a problem for self-supervised methods, if the parameters/hyperparameters of the methods are tuned for a specific sequence/anatomy as in our case, this could potentially impact the robustness of the methods, as these parameters/hyperparameters are obtained from training/tuning on 3D, brain MPRAGE scans acquired at 3T. This is despite the data consistency inherently embedded in CS-L1Wavelet, DeepDecoder, and SSDU. 

Generalizability and robustness of reconstruction methods have been studied in the context of end-to-end, supervised methods for MR reconstruction in \citep{knoll2019assessment,hammernik2021systematic,antun2020instabilities}. We briefly summarize some relevant conclusions from these articles. \citep{knoll2019assessment} found that that different domain shifts reduced performance more than others (e.g. changing SNR vs. image contrast), and that transfer learning is a viable strategy for handling distribution shifts. \citep{hammernik2021systematic} found that data consistency is important for robustness, and that at acceleration factor 4, distribution shifts are less of an issue. \citep{antun2020instabilities} found that supervised methods are vulnerable to adversarial perturbations, i.e. perturbations constructed such that minimal changes in the input data result in significant changes in the output. 

In \citep{robustCompressive}, the authors examine the robustness of end-to-end methods, compressed sensing, and variations of Deep Image Prior/DeepDecoder to distribution shifts, adversarial perturbations, and recovery of small features. They found that for both supervised and self-supervised methods, distribution shifts resulted in decreased PSNR/SSIM scores; in addition, the decrease was roughly the same for each method, preserving the ranking of the methods. Finally they found that all methods, including self-supervised methods, were vulnerable to adversarial attacks, including CS-L1Wavelet and DeepDecoder. Furthermore, \cite{zhang2021instabilities} showed the vulnerability of SSDU to adversarial attacks, showing that this was primarily due to the data consistency term. Thus, CG-SENSE can also be assumed to be vulnerable. Therefore, all the methods used in this paper have been shown to be vulnerable to adversarial attacks. We note that these works are based on retrospective reconstructions/retrospective sampling from fully-sampled datasets for their validation. 

In line with \citep{knoll2019assessment}, we found that different distribution shifts affected generalization differently; changing anatomy/contrast worsened the overall image quality rating in comparison to changing the field strength for all methods according to the MR scientists; in contrast, the radiologist found that only SSDU/CS-L1Wavelet worsened. However, as the mean scores in the knee scans for CG-SENSE/DeepDecoder were already 4 (the worst score), the decrease may not reflect any substantial difference in quality. As in \citep{hammernik2021systematic}, data consistency is crucial for the robustness of self-supervised methods as network parameters are trained solely through the modelling/the acquired undersampled data; in particular, we do not see any hallucination that can occur with end-to-end networks without data consistency. Furthermore, we see that as CG-SENSE produces a plausible image with acceleration factor 5, this can explain why distribution shifts were not so troublesome, as the self-supervised methods mainly needed to denoise, rather than recover anatomy/missing high frequency details. 

In contrast to \citep{robustCompressive}, our PSNR/SSIM results on the phantoms do not preserve the ranking between methods, although the PercDis results do, approximately. However, the qualitative metrics between distribution shifts over the different brain/knee scans seem to preserve ranking according to the no-reference image metrics/human ratings; this is consistent with PercDis being a better measure for perceptual image quality/similarity than PSNR/SSIM. In addition, the distribution shift in \citep{robustCompressive} was between two, similar datasets of knee MRI, as compared to our distribution shifts, where we change anatomy, contrast, etc.

For a clinical scenario, it was of interest to see if self-supervised methods could potentially work, without retraining, on other sequences, as retraining after deployment could be impractical. Furthermore, while adversarial perturbations are valuable for studying the input stability of reconstruction methods, they need to be manually constructed for each method and added to the input data. As clinical MR reconstruction is a closed loop, this kind of manual perturbation would require hacking the internal MR computer. Therefore, transfer learning and adversarial perturbations were outside the scope of this work, although from \citep{hammernik2021systematic,knoll2019assessment,robustCompressive}, we would expect an increase in image quality from transfer learning and vulnerability to adversarial perturbations for the methods considered in this paper. For example, \citep{darestani2021accelerated} found, in a retrospective study, that DeepDecoder had different optimal (judged by PSNR/SSIM) hyperparameters for brain vs. knee scans. However, SSDU and CS-L1Wavelet, tuned only on 3T MPRAGE brain data, are able to achieve an overall image quality of fair to good on a diverse dataset. 

\subsection{Ranking Methods and Quantitative Metrics}
From a perceptual viewpoint (PercDis score, no-reference image metrics, human rating), SSDU and CS-L1Wavelet performed the best, with an edge to SSDU in the PercDis score/no-reference image metrics. From a pixel-wise metric viewpoint (PSNR,SSIM), DeepDecoder was better than or similar to all methods, as was also found in \citep{robustCompressive}. CG-SENSE consistently performed the worst or similarly to all methods over all metrics. With respect to validation, both approaches have their advantages and disadvantage; while pixel-wise metrics are the natural way to compare against a ground-truth, they may not correlate well with the perception of a radiologist. While perceptual metrics may be intuitive, the absence of ground truth can make it less objective. To our knowledge, current state of the art MR image reconstructions are generally not evaluated with perceptual metrics such as PercDis or \citep{adamson2021ssfd}, which require ground truth, or the no-reference image quality metrics. However, given the close correspondence of the image quality metrics/PercDis to the human ratings/perceptual evaluation, as well as other evidence from the literature \citep{woodard2006no,zhang2018can}, perceptual metrics could be used as a complement to pixel-wise metrics/human ratings. 

\subsection{Implications for Future Methods and Validation}
We note that while SSDU generally outperformed DeepDecoder, SSDU's denoising network was trained on a dataset of 3T MPRAGE, thus learning a prior over multiple subjects. In contrast, DeepDecoder only learns/performs inference over a single slice at a time, thus limiting the amount of information in comparison to SSDU. In \cite{korkmaz2022unsupervised}, the authors show that a Deep Image Prior based reconstruction can be fused explicitly with prior information from a dataset of fully sampled acquistions to increase performance; such fusion of prior information could potentially also benefit DeepDecoder and other self-supervised methods which operate on a per slice basis. In addition, from the qualitative results, CS-L1Wavelet with regularization parameter tuned using the Noise2Self framework is competitive with SSDU. At lower acceleration factors, such as the one used in this paper, it is plausible that this result generalizes, such that compressed sensing reconstructions with optimally tuned regularization parameters can be competitive with state of the art machine learning methods, at least on a qualitative basis. For future validation, we conclude that appropriate regularization parameter tuning strategies should be used when comparing compressed sensing reconstructions to new methods. Finally, we note that as the theory behind the self-supervised methods we used (Deep Image Prior and blind denoising) form the basis for or are conceptually similar to many other self-supervised methods, it is plausible that the impressive robustness showed by these methods to a diverse range of realistic distribution shifts would generalize to future self-supervised methods.

\subsection{Future of Validation}
However, whatever metrics or datasets are used for validating methods, the ultimate test for reconstruction methods is the usefulness to radiologists for reliably diagnosing pathology in comparison to currently used methods \citep{recht2020using,roux2019mri}. This can imply many things, including fine grained analysis of small textures/details/pathologies as well as tissue specific analysis, requiring novel datasets with extensive annotations by radiologists. \citep{zhao2021fastmri,desai2021skm} are two recent works in this direction, providing datasets with bounding box annotations/pathology annotations to further validate reconstructions. To assist validating future methods, the datasets acquired for this paper will be made available online; see \url{https://www.melba-journal.org/papers/2022:022.html} for details.

\section{Conclusion}

Rigorous validation is required to introduce new reconstruction algorithms into clinical routines. In this study, validation of prospective reconstructions, generalizability, and different image quality metrics were investigated. The results show that self-supervised image reconstruction methods have potential, but that further development is required to not only improve image quality but also to define a reliable, standardized way of validating new methods. Reliable validation can facilitate quicker translation to the clinical routine, with the ultimate goal of improving patient care.


\acks{This project is supported by the European Union’s Horizon 2020 research and innovation programme under the Marie Sklodowska-Curie project TRABIT (agreement No 765148 to TY) and by the Swiss National Science Foundation (SNSF, Ambizione grant PZ00P2\_185814 to EJC-R) We acknowledge access to the facilities and expertise of the CIBM Center for Biomedical Imaging, a Swiss research center of excellence founded and supported by Lausanne University Hospital (CHUV), University of Lausanne (UNIL), Ecole Polytechnique Fédérale de Lausanne (EPFL), University of Geneva (UNIGE) and Geneva University Hospitals (HUG). 
}

%
\ethics{The work follows appropriate ethical standards in conducting research and writing the manuscript, following all applicable laws and regulations regarding treatment of animals or human subjects.}

\coi{Thomas Yu, Gian Franco Piredda, Gabriele Bonanno, Arun Joseph, Tom Hilbert and Tobias Kober are employed by Siemens Healthineers International AG, Switzerland. }

\appendix 
\section*{Appendix}
\subsection*{Hyperparameter Tuning}
For example, to set the regularization parameter of CS-L1Wavelet, we treat it as a function with a single parameter ($\lambda$). We can then optimize this parameter using the Noise2Self training framework to estimate the $\lambda$ which minimizes the noise-free error between simulated measurements and the acquired measurements. Concretely, we fix 20 logarithmically spaced values from 0.00001 to 0.1. We set each value as $\lambda$ and run 50 image reconstructions corresponding to different, random masks and average the corresponding errors with respect to the complementary mask in order to approximate the true measurement error associated with using each value. We then select the value with the lowest measurement error as the optimal regularization parameter. This is done for each slice in each subject; the final regularization value which is used throughout this paper is the average over all subjects. The hyperparameters of DeepDecoder and SSDU are set similarly with a grid search over the network hyperparameters, albeit over a much smaller set of data due to the high computational demand. 
\begin{landscape}

\begin{table}
\resizebox{1.4\textwidth}{!}{
\begin{tabular}{lllllllllll}
                                                  & \textbf{1.} & \textbf{2.} & \textbf{3.} & \textbf{4.} & \textbf{5.} & \textbf{6.} & \textbf{7.} & \textbf{8.} & \textbf{9.} & \textbf{10.} \\ \cline{2-11} 
\multicolumn{1}{l|}{\textbf{Sequence Type}}       & MPRAGE      & MPRAGE      & MPRAGE      & MPRAGE      & MPRAGE      & MPRAGE      & SPACE       & SPACE       & SPACE       & SPACE        \\
\multicolumn{1}{l|}{\textbf{Field Strength (T)}}  & 1.5         & 3           & 7           & 3           & 3           & 3           & 3           & 3           & 3           & 3            \\
\multicolumn{1}{l|}{\textbf{Body Part}}           & Brain       & Brain       & Brain       & Brain       & Brain       & Brain       & Brain       & Brain       & Knee        & Knee         \\
\multicolumn{1}{l|}{\textbf{Coils}}               & 1Tx/20Rx    & 1Tx/64Rx    & 8pTx/32Rx   & 1Tx/20Rx    & 1Tx/64Rx    & 1Tx/64Rx    & 1Tx/64Rx    & 1Tx/64Rx    & 1Tx/18Rx    & 1Tx/18Rx     \\
\multicolumn{1}{l|}{\textbf{Resolution (mm\textsuperscript{3})}}    & 1.3x1.3x1.2 & 1x1x1       & 0.7x0.7x0.7 & 1x1x1       & 1x1x1       & 1x1x1       & 1x1x1       & 1x1x1       & 0.3x0.3x0.6 & 0.3x0.3x0.6  \\
\multicolumn{1}{l|}{\textbf{Field of View (mm\textsuperscript{3})}} & 240x240x160 & 256x240x208 & 250x219x179 & 256x240x208 & 256x240x208 & 256x240x208 & 250x250x176 & 250x250x176 & 160x160x134 & 160x160x115  \\
\multicolumn{1}{l|}{\textbf{Inversion Time (s)}}  & 1           & 0.9         & 1.1         & 0.9         & 0.9         & 0.972       & -           & 2.05        & -           & -            \\
\multicolumn{1}{l|}{\textbf{Repetition Time (s)}} & 2.4         & 2.3         & 2.5         & 2.3         & 2.3         & 1.93        & 0.7         & 7           & 0.9         & 1            \\
\multicolumn{1}{l|}{\textbf{Echo Time (ms)}}      & 3.47        & 2.9         & 2.87        & 2.9         & 2.9         & 2.61        & 11          & 392         & 29          & 108          \\
\multicolumn{1}{l|}{\textbf{Echo Spacing (ms)}}   & 7.86        & 6.88        & 7.8         & 6.88        & 6.88        & 6.28        & 3.72        & 3.66        & 4.84        & 5.12         \\
\multicolumn{1}{l|}{\textbf{Bandwidth (Hz/Px)}}   & 180         & 240         & 250         & 240         & 240         & 280         & 630         & 651         & 488         & 416          \\
\multicolumn{1}{l|}{\textbf{Turbo Factor}}        & 192         & 198         & 250         & 198         & 198         & 198         & 42          & 220         & 35          & 44           \\
\multicolumn{1}{l|}{\textbf{Acceleration Factor}} & 4.2         & 5           & 5           & 5           & 5           & 5           & 4           & 6           & 7           & 7            \\
\multicolumn{1}{l|}{\textbf{Acquisition Time}}    & 1:28 min    & 1:34 min    & 2:42 min    & 1:34 min    & 1:34 min    & 1:20 min    & 3:27 min    & 3:46 min    & 4:41 min    & 3:52 min    
\end{tabular}
}
\caption[Sequence Parameters]{ Detailed sequence parameters of all used datasets. We note that spiral phyllotaxis sampling \cite{mussard2020accelerated} and Poisson disk sampling \cite{bridson2007fast} were used for the MPRAGE and SPACE sequences respectively.}
\label{tab:seqparam}
\end{table}

\end{landscape}
\newpage
\begin{figure} 
\includegraphics[width=\textwidth]{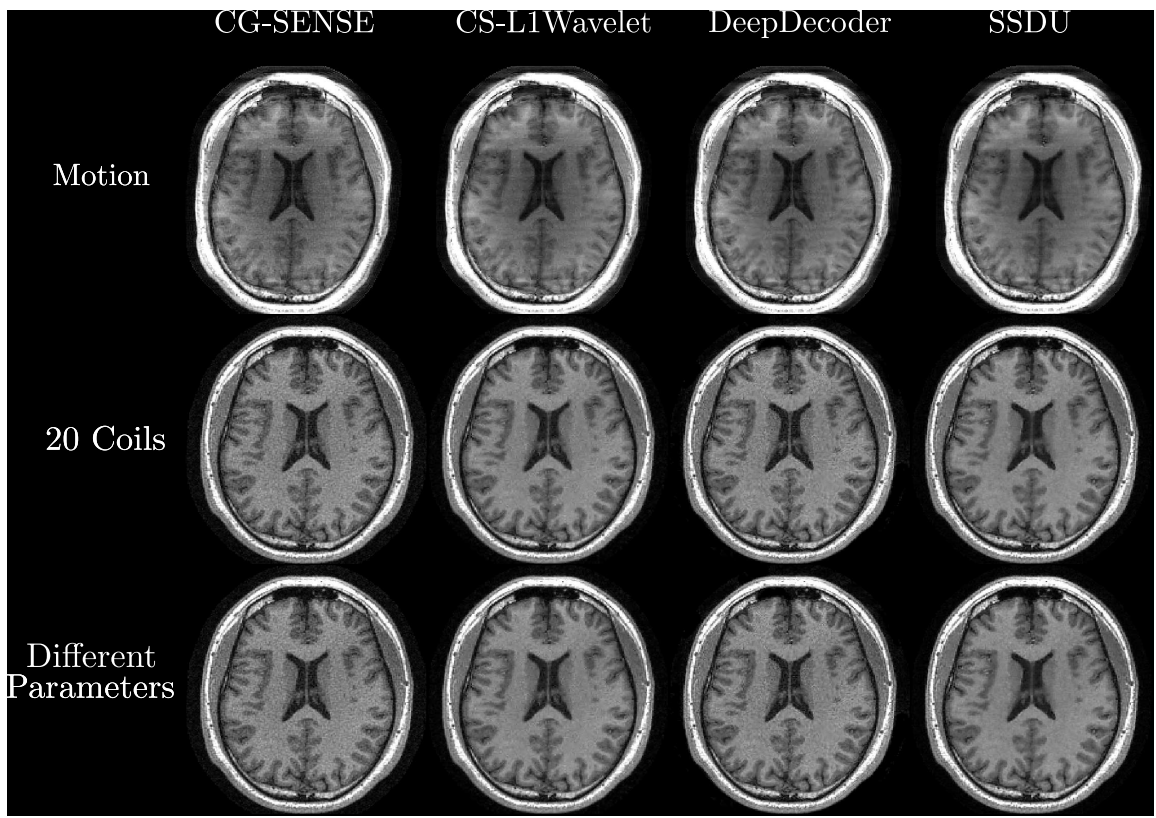}
\caption[MPRAGE: Axial brain slices from different perturbations]{Here we show axial brain slice reconstructions from three different perturbations of the MPRAGE sequence: the addition of motion, using 20 coils instead 64 coils, and changing the parameters of the MPRAGE sequence. The images are \textbf{not} registered due to interpolation effects from co-registration. }
\label{S-fig:mprage_variations}
\end{figure}

\begin{figure} 
\includegraphics[width=\textwidth]{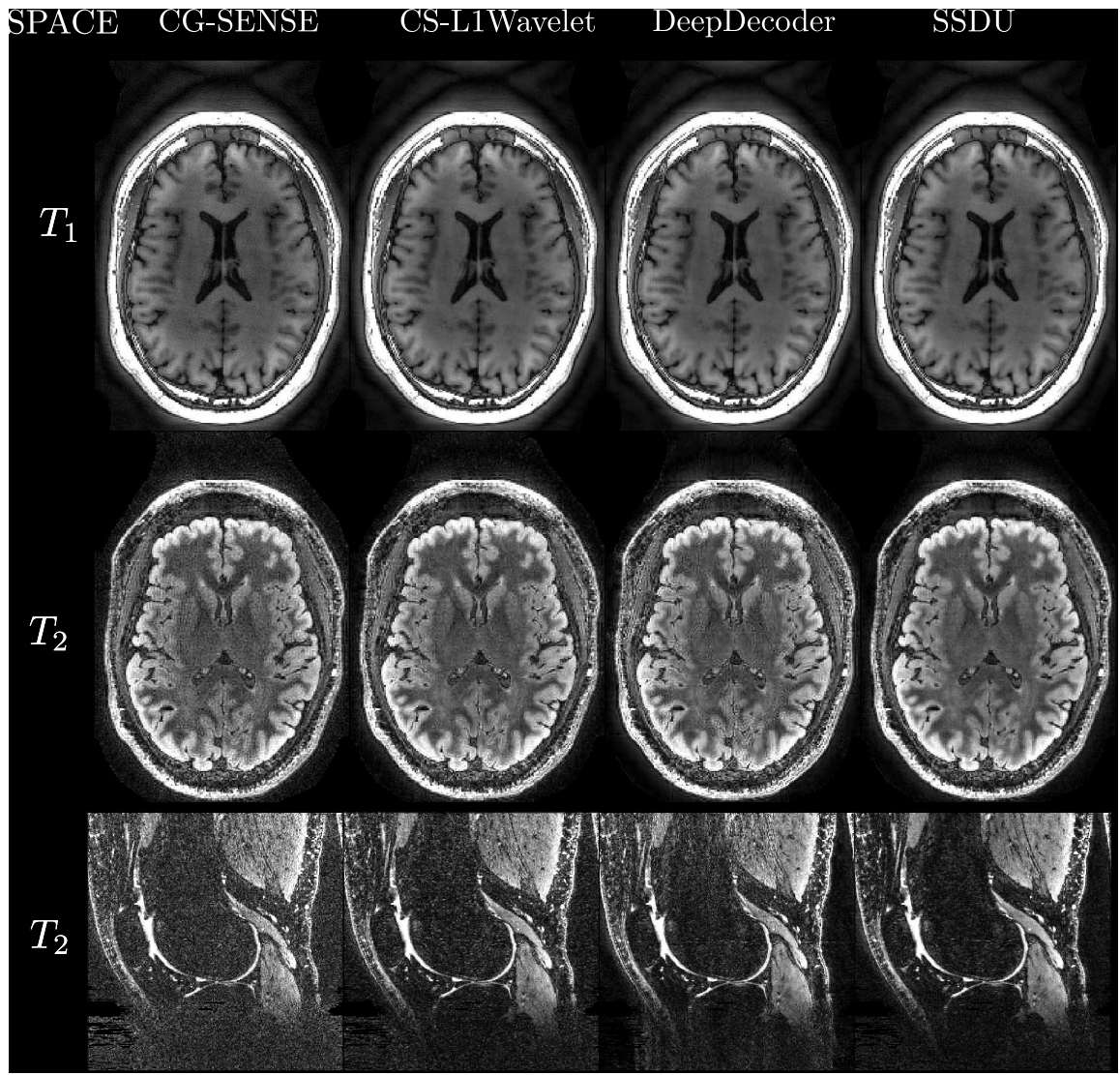}
\caption[SPACE: Axial brain slices and a sagittal knee slice]{Here we show axial brain slices and a sagittal knee slice from the  reconstructions from the SPACE acquisitions. The images are \textbf{not} registered due to interpolation effects from co-registration.  }
\label{S-fig:space_variations}
\end{figure}

\end{document}